\documentclass[english,aps,prl,twocolumn,superscriptaddress]{revtex4}
\usepackage[T1]{fontenc}
\usepackage[latin9]{inputenc}
\usepackage{amstext}
\usepackage{graphicx}
\usepackage{amssymb}
\usepackage{multirow}
\usepackage{dcolumn}
\usepackage{color}
\usepackage{amsmath}
\usepackage{babel}
\usepackage{epstopdf}
\usepackage{natbib}
\makeatletter

\begin{document}

\title{Measuring and Suppressing Quantum State Leakage in a Superconducting Qubit}

\author{Zijun Chen}
\affiliation{Department of Physics, University of California, Santa Barbara, California 93106-9530, USA}
\author{Julian Kelly}
\affiliation{Google Inc., Santa Barbara, California 93117, USA}
\author{Chris Quintana}
\affiliation{Department of Physics, University of California, Santa Barbara, California 93106-9530, USA}
\author{R. Barends}
\affiliation{Google Inc., Santa Barbara, California 93117, USA}
\author{B. Campbell}
\affiliation{Department of Physics, University of California, Santa Barbara, California 93106-9530, USA}
\author{Yu Chen}
\affiliation{Google Inc., Santa Barbara, California 93117, USA}
\author{B. Chiaro}
\affiliation{Department of Physics, University of California, Santa Barbara, California 93106-9530, USA}
\author{A. Dunsworth}
\affiliation{Department of Physics, University of California, Santa Barbara, California 93106-9530, USA}
\author{A. Fowler}
\affiliation{Google Inc., Santa Barbara, California 93117, USA}
\author{E. Lucero}
\affiliation{Google Inc., Santa Barbara, California 93117, USA}
\author{E. Jeffrey}
\affiliation{Google Inc., Santa Barbara, California 93117, USA}
\author{A. Megrant}
\affiliation{Department of Physics, University of California, Santa Barbara, California 93106-9530, USA}
\affiliation{Department of Materials, University of California, Santa Barbara, California 93106, USA}
\author{J. Mutus}
\affiliation{Google Inc., Santa Barbara, California 93117, USA}
\author{M. Neeley}
\affiliation{Google Inc., Santa Barbara, California 93117, USA}
\author{C. Neill}
\affiliation{Department of Physics, University of California, Santa Barbara, California 93106-9530, USA}
\author{P. J. J. O'Malley}
\affiliation{Department of Physics, University of California, Santa Barbara, California 93106-9530, USA}
\author{P. Roushan}
\affiliation{Google Inc., Santa Barbara, California 93117, USA}
\author{D. Sank}
\affiliation{Google Inc., Santa Barbara, California 93117, USA}
\author{A. Vainsencher}
\affiliation{Department of Physics, University of California, Santa Barbara, California 93106-9530, USA}
\author{J. Wenner}
\affiliation{Department of Physics, University of California, Santa Barbara, California 93106-9530, USA}
\author{T. C. White}
\affiliation{Department of Physics, University of California, Santa Barbara, California 93106-9530, USA}

\author{A. N. Korotkov}
\affiliation{Department of Electrical and Computer Engineering, University of California, Riverside, California 92521, USA}
\author{John M. Martinis}
\email{jmartinis@google.com}
\affiliation{Department of Physics, University of California, Santa Barbara, California 93106-9530, USA}
\affiliation{Google Inc., Santa Barbara, California 93117, USA}
\date{\today}
\begin{abstract}
Leakage errors occur when a quantum system leaves the two-level qubit subspace. Reducing these errors is critically important for quantum error correction to be viable. To quantify leakage errors, we use randomized benchmarking in conjunction with measurement of the leakage population. We characterize single qubit gates in a superconducting qubit, and by refining our use of Derivative Reduction by Adiabatic Gate (DRAG) pulse shaping along with detuning of the pulses, we obtain gate errors consistently below $10^{-3}$ and leakage rates at the $10^{-5}$ level. With the control optimized, we find that a significant portion of the remaining leakage is due to incoherent heating of the qubit.
\end{abstract}
\maketitle

Accurate manipulation of the states in a quantum two-level system (qubit) is a key requirement for building a fault tolerant quantum processor \cite{divincenzo2000physical}. However, many physical quantum systems such as quantum dots \cite{schliemann2001double} and superconducting qubits \cite{clarke2008superconducting} have multiple levels, from which two levels are chosen to form the computational subspace. The presence of non-computational levels leads to two types of errors: leakage errors where the quantum state populates non-computational levels, and phase errors due to coupling of computational and non-computational levels when driven by control fields \cite{martinis2014fast, gambetta2011analytic}. Previous experimental work \cite{lucero2010reduced, chow2010optimized} on superconducting qubits has focused on reducing phase errors, because they were the dominant source of total gate infidelity. Indeed, the suppression of phase errors using Derivative Reduction by Adiabatic Gate (DRAG) pulse shaping \cite{motzoi2009simple} has helped push single qubit fidelity in superconducting qubits over 99.9\%, nominally satisfying one of the requirements for realizing quantum error correction (QEC)  \cite{divincenzo2009fault, fowler2012surface}.

However, total fidelity is not the only metric that determines the viability of QEC because certain errors are more deleterious than others. Specifically, leakage errors are known to be highly detrimental for error correcting codes such as the surface code, because interactions with a qubit in a leakage state have a randomizing effect on the interacting qubits \cite{fowler2013coping}. Moreover, leakage states can be as long-lived as the qubit states, leading to time-correlated errors which further degrade performance \cite{ghosh2013understanding}. These concepts were recently demonstrated in a 9 qubit repetition code \cite{kelly2015state}, where single leakage events persisted for multiple error detection cycles and propagated errors to neighboring qubits. Understanding and reducing leakage is of critical importance for realizing QEC.
 
In this Letter, we characterize single qubit leakage errors in a superconducting qubit.  To estimate leakage errors, we use randomized benchmarking (RB) \cite{knill2008randomized,magesan2011scalable} in conjunction with measurements of leakage state populations. Using this method, we show that previous experimental realizations of DRAG pulse shaping have a tradeoff between total fidelity and leakage errors. We overcome this tradeoff using additional pulse shaping, and obtain gates that have both state of the art fidelity and low leakage. Additionally, we use RB to measure the dependence of leakage on pulse length.

Our experiment uses Clifford based randomized benchmarking \cite{magesan2011scalable}, which is typically used to characterize overall gate fidelity. In Clifford based RB, we apply a random sequence of gates chosen from the single qubit Clifford group, which is the group of rotations that map the six axial Bloch states to each other. We then append a recovery Clifford gate to the end of the sequence such that the complete sequence is ideally the identity operation. Thus, the fidelity of a sequence is the probability of mapping $|0\rangle$ to $|0\rangle$. By randomly choosing the gates in each sequence, phase and amplitude errors accumulate incoherently, which leads to exponential decay of the sequence fidelity with increasing sequence length. The crux of our protocol is that randomization also accumulates leakage errors incoherently \cite{supplement}, such that over many gates we build up leakage populations to a level that can be measured using current techniques. We note that leakage errors as discussed here differ from irreversible loss of the qubit; RB in the presence of loss was previously discussed in Ref.\,\cite{wallman2015robust}.

For our testbed we use a single Xmon transmon qubit
 \cite{barends2013coherent,koch2007charge}  ($Q_7$) from the 9 qubit chain described in Ref.\,\cite{kelly2015state}. The transmon has a weakly anharmonic potential, shown in Fig.\,1(a), which supports a ladder of energy levels. The two lowest levels form our qubit, and the primary non-computational level is the $|2\rangle$ state. Leakage errors arise when the qubit state is directly excited to the $|2\rangle$ state, while phase errors occur due to AC Stark shifting of the 1$\leftrightarrow$2 transition \cite{martinis2014fast}. 

We operate the qubit at a frequency $f_{10}$ of around 5.3\,\,GHz, and the anharmonicity $\Delta = \omega_{21} - \omega_{10}$ is $2\pi \times -212$\,MHz. Microwave (XY) control is achieved using a capacitively coupled transmission line driven at the qubit frequency. We generate control pulses using a custom arbitrary waveform generator, and the pulses are shaped with a cosine envelope. We measure the qubit state using a dispersive readout scheme \cite{wallraff2005approaching} in conjunction with a bandpass filter \cite{jeffrey2014fast} and a wideband parametric amplifier \cite{mutus2014strong}. This setup allows us to discriminate the $|2\rangle$ state in addition to the two computational levels with high fidelity \cite{supplement}.  The $T_1$ of the device at the operating frequency is 22\,\,$\mu$s, while a Ramsey experiment shows two characteristic decay times  \cite{rbidles}, an exponential decay time $T_{\phi_1}$ of 8\,$\mu$s and a Gaussian decay time $T_{\phi_2}$ of 1.8\,$\mu$s.

To illustrate our novel use of RB, we begin by measuring how DRAG suppresses leakage and phase errors. We use the simplified version of DRAG described in Refs.\,\cite{lucero2010reduced, chow2010optimized}. Given a control envelope $\Omega(t)$, we add the time derivative $\dot{\Omega}(t)$ to the quadrature component:
\begin{equation}
\Omega'(t) = \Omega(t) - i \alpha \frac{\dot{\Omega}(t)}{\Delta}
\end{equation}
where $\alpha$ is a weighting parameter. Fourier analysis \cite{martinis2014fast, motzoi2013improving} shows that the DRAG correction suppresses the spectral weight of the control pulse at the $1\leftrightarrow2$ transition if $\alpha=1.0$, which minimizes leakage errors. However, the optimal value to compensate the AC Stark shift and correct for phase errors is $\alpha=0.5$  \cite{martinis2014fast, lucero2010reduced}.

\begin{figure}
\begin{centering}
\includegraphics[width=3.25 in]{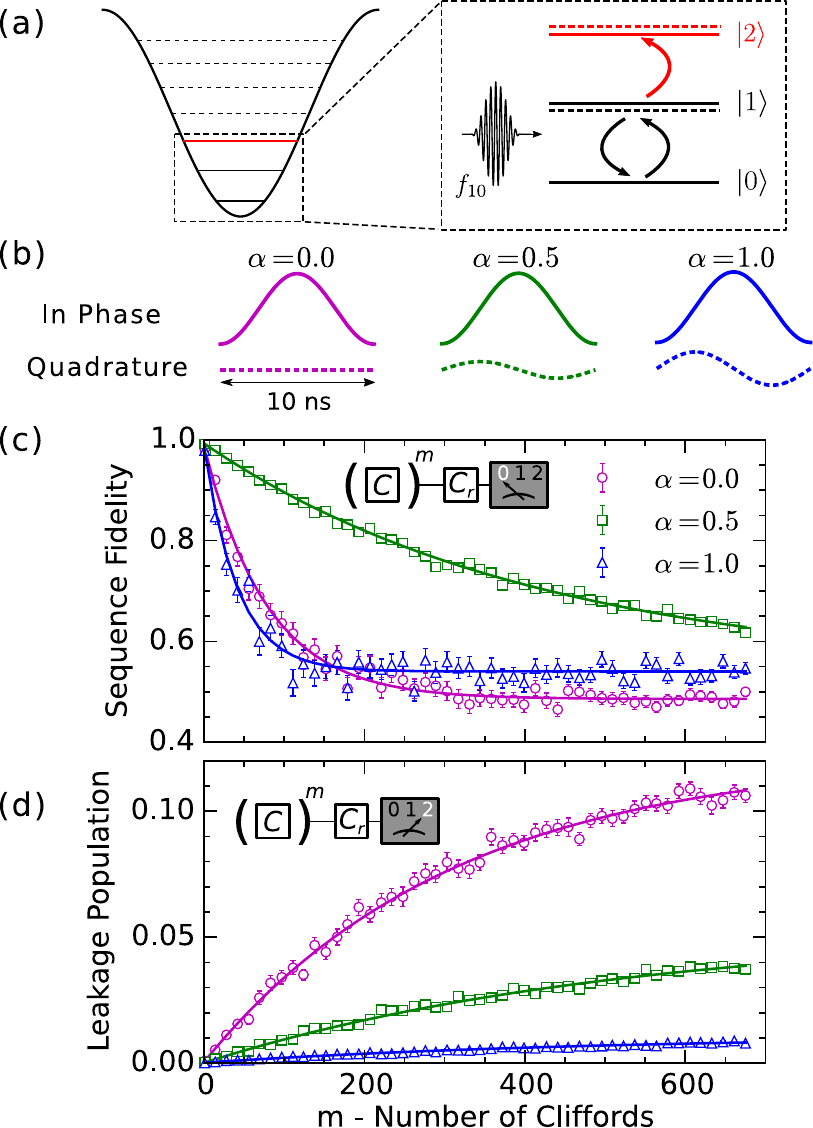} 
\par\end{centering}
\caption{(a) Weakly anharmonic potential of a transmon. When driving $|0\rangle$ to $|1\rangle$, direct excitation to $|2\rangle$ (red arrow) causes leakage errors, while AC Stark repulsion of the 1$\leftrightarrow$2 transition (dashed lines) leads to phase errors. (b) The simple DRAG correction, which adds the derivative of the envelope to the quadrature component of the envelope. Three different DRAG weightings ($\alpha$) are shown. (c) Exponential decay of sequence fidelity from randomized benchmarking, showing data for the three values of $\alpha$. Each point is the average of 75 different random sequences. Fidelity is highest for $\alpha=0.5$ (d) $|2\rangle$ state population \textit{vs} sequence length, showing the accumulation of leakage with sequence length. Leakage is lowest for $\alpha=1.0$.}
\label{figure:rbexamples} 
\end{figure}

We confirm these concepts by performing Clifford based RB using 10\,ns microwave pulses shaped with three different values of $\alpha$ (0, 0.5, and 1.0), as shown in Fig.\,1(b). We combine up to three pulses to form a single Clifford gate; on average each Clifford contains 1.5 $\pi/2$-pulses and 0.375 $\pi$-pulses, resulting in an average gate length of 18.75\,ns. Figure 1(c) shows  sequence fidelity decay curves for the three values of $\alpha$. As expected, using $\alpha=0.5$ results in higher fidelities than $\alpha = 0.0$ or $\alpha = 1.0$. We can quantify this improvement from the characteristic scale of the decay $p$, obtained by fitting to $A p^m + B$ where $A$ and $B$ encapsulate state preparation and measurement errors. We then estimate the error per Clifford as $r_{\text{Clifford}}=(1-p)/2$ \cite{magesan2011scalable}. For $\alpha=0.5$, we obtain an error per Clifford of $9.6 \pm 0.1 \times 10^{-4}$, while for $\alpha=0.0$ and $\alpha=1.0$ we obtain errors of $6.3 \pm 0.2 \times 10^{-3}$ and $1.20 \pm 0.01 \times 10^{-2}$ per Clifford, respectively.

Simultaneously, we characterize leakage errors in our gateset from the dynamics of the $|2\rangle$ state measured while performing RB, as shown in Fig.\,1(d). For all three value of $\alpha$, the $|2\rangle$ state population shows an exponential approach to a saturation population. Without correction, this saturation population is significant at about 10\%, but decreases by about a factor of 3 for $\alpha=0.5$ and by a factor of 10 for $\alpha=1.0$. To quantify the leakage rate per Clifford, we fit the $|2\rangle$ state dynamics to a simple rate equation that takes into account leakage from the computational subspace into the $|2\rangle$ state and decay of the $|2\rangle$ state back into the subspace \cite{supplement}.

\begin{align}
p_{|2\rangle}(m) &= p_\infty \left(1 - e^{-\Gamma m} \right) + p_0 e^{-\Gamma m} \\
\Gamma &= \gamma_\uparrow + \gamma_\downarrow \,\,\,\,\,\,\,\, p_\infty =\gamma_\uparrow / \Gamma
\end{align}
where $p_{|2\rangle}(m)$ is the $|2\rangle$ state population as a function of sequence length $m$, $\gamma_\uparrow$ and $\gamma_\downarrow$ are the leakage and decay rates per Clifford, and $p_0$ is the initial $|2\rangle$ state population. Using Eq.\,(2), we extract leakage rates of $3.92 \pm 0.08 \times 10^{-4}$, $1.02 \pm 0.02 \times 10^{-4}$, and $2.18 \pm 0.08 \times 10^{-5}$ for $\alpha = $0, 0.5 and 1.0.

The results from RB confirm the theory behind simple DRAG: we can minimize either phase error or leakage error, but not both. To simultaneously optimize for both gate fidelity and leakage performance, we would like to minimize leakage using simple DRAG, then separately compensate the AC Stark shift. In the original DRAG theory, the Stark shift was compensated using a time dependent detuning of the qubit \cite{motzoi2009simple}. However, as was previously noted in Refs.\,\cite{martinis2014fast, gambetta2011analytic}, a constant detuning should also be able to compensate the AC Stark shift. Given an envelope $\Omega'$, which in general can have a quadrature correction, we generate a new envelope
\begin{equation}
\Omega''(t) = \Omega'(t) e^{2 \pi i \, \delta f \, t}
\end{equation}
where $\delta f$ is the detuning of the pulse from the qubit frequency.  We also redefine the anharmonicity parameter in Eq.\,(1) to be $\Delta = \omega_{21} - (\omega_{10}+2\pi\,  \delta f)$, so that leakage suppression still occurs at the $1 \leftrightarrow 2$ frequency. An example of a detuned pulse is shown in Fig.\,2(a).

\begin{figure}
\begin{centering}
\includegraphics[width=3.25 in]{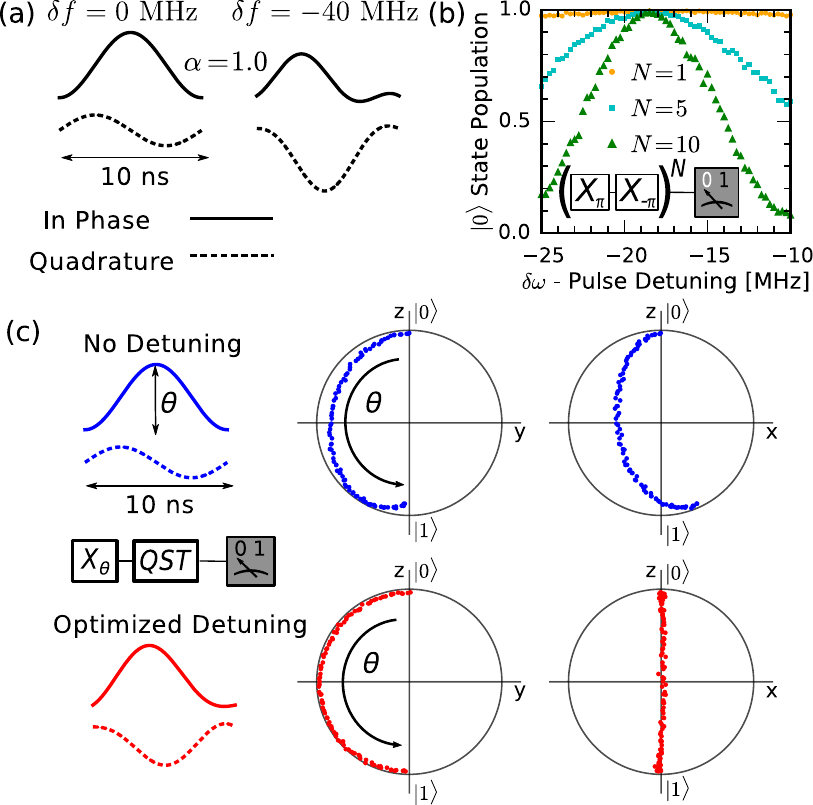} 
\par\end{centering}
\caption{(a) Control envelopes with simple DRAG with (right) and without (left) detuning of the pulse. The detuning is exaggerated for illustration. (b) We sweep over the detuning $\delta f$ while performing the pseudo-identity sequence shown in the inset. The sequence maps back to $|0\rangle$ when the detuning is optimized. Multiple applications of the pulse sequence increases the sensitivity of the measurement. (c) Quantum state trajectories plotted on projections of the Bloch sphere, with (bottom) and without (top) optimal detuning. The data is obtained by performing quantum state tomography (QST) after applying a variable X rotation, with the rotation angle ranging from 0 to $\pi$.}
\label{figure:detuning} 
\end{figure}

To optimize the detuning parameter $\delta f$, we sweep the detuning of a $\pi$-pulse while performing the psuedo-identity operation of a $\pi$-pulse followed by a $-\pi$-pulse along the same rotation axis \cite{lucero2010reduced, pulsedistortion}. As shown in Fig.\,2(b), the pulse detuning is optimized when the $|0\rangle$ state population is maximized, and the psuedo-identity can be applied multiple times to increase the resolution of the measurement. To verify that the detuning has suppressed phase errors, we perform quantum state tomography after applying a control pulse to our qubit while ramping the amplitude of the pulse, as shown in Fig.\,2(c). Without detuning, the Bloch vector never reaches the pole, while the behavior is much closer to ideal when the detuning is optimized.
\begin{figure}
\begin{centering}
\includegraphics[width=3.25 in]{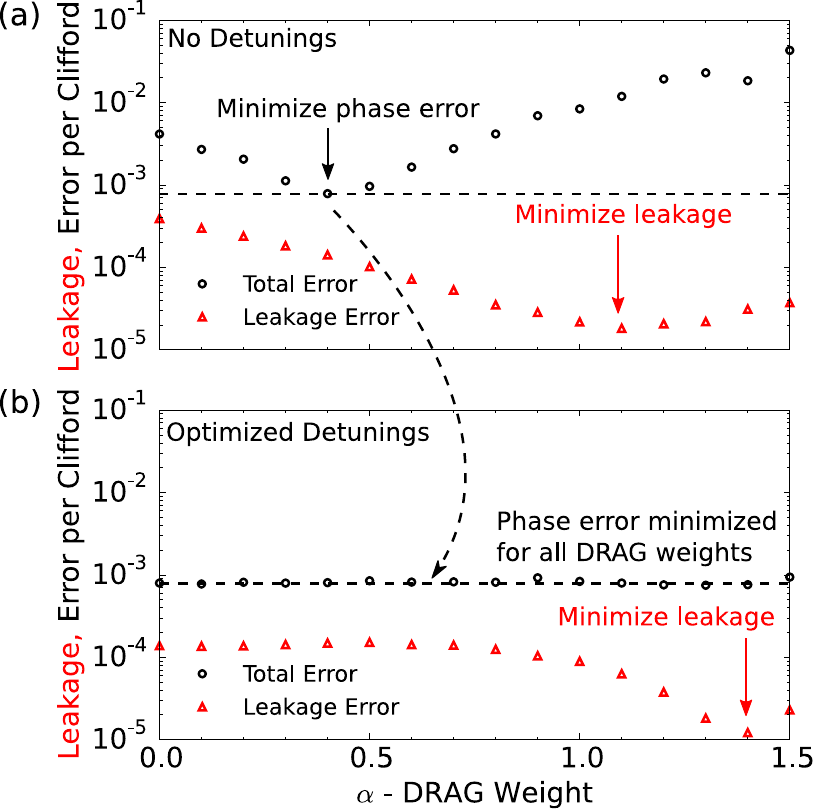} 
\par\end{centering}
\caption{Total gate fidelity and leakage rates versus DRAG weighting $\alpha$, measured using RB. (a) Without using pulse detunings, we require different values of $\alpha$ to minimize overall error versus leakage errors. (b) By optimizing our pulses using detunings, we obtain high fidelity for any $\alpha$, and are free to choose $\alpha$ to minimize leakage.}
\label{figure:rbvdrag} 
\end{figure}

We now explore in more detail the dependence of fidelity and leakage on $\alpha$. In Fig.\,3, we show parameters extracted from RB with 10\,ns pulses while varying $\alpha$ between 0.0 and 1.5. Without detuning the pulses, we find the minimum error per Clifford to be $7.9 \pm 3 \times 10^{-4}$ when $\alpha=0.4$. We note that this is a deviation from the expected optimal value of $\alpha=0.5$; we attribute this deviation to distortions of the pulse between the waveform generator and the qubit \cite{pulsedistortion}. Away from the optimal $\alpha$, the error increases rapidly. 

Next, we optimize the detuning of the pulses for each value of $\alpha$ using the method described in Fig.\,2. We find that when using $\pi$ and $\pi/2$ pulses with the same length, using the same detuning for both types of pulses yields the best results. After calibrating the detuning, we recalibrate the pulse amplitudes, then run a short Nelder-Mead optimization on the RB fidelity to get final adjustments to pulse parameters \cite{kelly2014optimal}. With these optimizations, we find that the average error per Clifford for all values of $\alpha$ to be $9.1 \times 10^{-4}$, with a standard deviation of $1 \times 10^{-4}$. In other words, we can tune up high fidelity gates for \emph{any} value of $\alpha$.

With gate fidelity now independent of $\alpha$, we are free to implement DRAG solely to minimize leakage. Without detuning, the minimum leakage rate is $1.82 \pm 0.07 \times 10^{-5}$ for $\alpha=1.1$. After detuning the pulses for optimal fidelity, we see shifts in the leakage rates. For $\alpha > 0.4$, we detune the pulses towards the 1$\leftrightarrow$2 transition \cite{supplement} which tends to increase the leakage rate. Nevertheless, we can still suppress leakage to the same level as the undetuned pulses by increasing $\alpha$ to $1.4$. Using these parameters, we achieve both high fidelity  ($8.7 \pm 0.4 \times 10^{-4}$ error per Clifford) and low leakage ($1.2 \pm 0.1 \times 10^{-5}$) \cite{supplement}. 
\begin{figure}
\begin{centering}
\includegraphics[width=3.25 in]{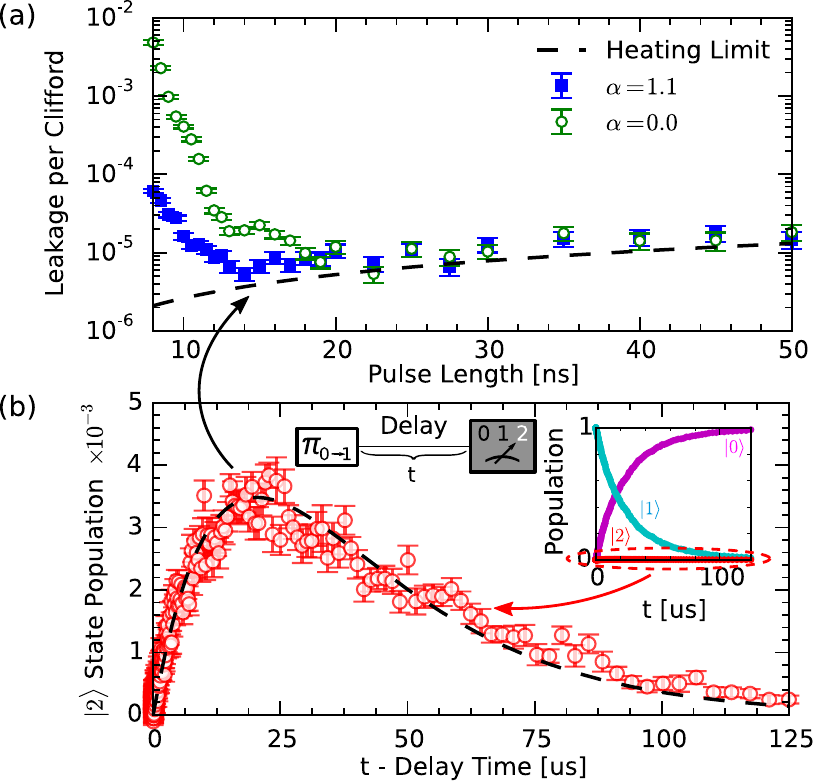} 
\par\end{centering}
\caption{(a) Leakage rate per Clifford extracted from RB versus pulse length, with $\alpha=0.0$ and $\alpha=1.1$. The dashed line is the lower bound on leakage calculated from the heating rate. (b) Heating of the qubit from $|1\rangle$ to $|2\rangle$. We prepare the qubit in $|1\rangle$, wait for time $t$, then measure the qubit state. Inset: The dynamics of all three states, primarily showing $T_1$ decay of $|1\rangle$ to $|0\rangle$. Main figure: Zoom in of the $|2\rangle$ state dynamics, showing an increase in population due to heating before relaxing back to zero. The data has been corrected for readout visibility. The dashed line is a rate equation fit, from which we extract the heating rate plotted in (a).}
\label{figure:leakvlength} 
\end{figure}

Having characterized 10\,ns pulses in detail, we now examine the dependence of leakage on pulse length.  As noted in Fig.\,3, pulse detuning can affect the leakage rate; for simplicity we set the detuning to zero for the following measurements. We initially set $\alpha=0.0$ and measure the leakage rate while varying the length of our pulses between 8\,ns and 50\,ns and calibrating the pulse amplitudes accordingly. We then repeat this measurement with $\alpha=1.1$ where we previously found leakage to be suppressed in Fig.\,3(a). The results are shown in Fig.\,4(a). For short pulses, we observe that the leakage rate decreases exponentially with increasing pulse length, and that the DRAG correction generally suppresses leakage by an order of magnitude or more. However, as the pulse length increases past 15\,ns, the leakage rate begins to level off and even begin to increase. Furthermore, the effect of DRAG is no longer distinguishable for pulses longer than 20\,ns. These results suggest that for long pulses, leakage is the result of incoherent processes such as thermal excitations or noise at the 1$\leftrightarrow$2 transition, rather than coherent processes such as control errors.

To measure the incoherent leakage rate, we prepare the qubit in the $|1\rangle$ state and measure the dynamics of the three qubit states, as shown in Fig.\,4(b). We see that the $|2\rangle$ state population initially rises over 20\,$\mu$s, corresponding to heating from $|1\rangle$ to $|2\rangle$. Then, the $|2\rangle$ population slowly decays to zero as both excited states relax due to $T_1$ processes. We model the $|2\rangle$ population using a rate equation with three rates: decay from  $|2\rangle$ to $|1\rangle$, decay from $|1\rangle$ to $|0\rangle$, and heating from $|1\rangle$ to $|2\rangle$. We ignore nonsequential transitions since they are suppressed in the nearly harmonic transmon potential \cite{peterer2015coherence}, as well as heating from $|0\rangle$ to $|1\rangle$ since we assume the initial state is $|1\rangle$. We extract the two decay rates from $T_1$ measurements, which give $T_1^{|1\rangle}=22$\,$\mu$s and $T_1^{|2\rangle}=18$\,$\mu$s \cite{footnote}. The remaining parameter to fit is the $1 \rightarrow 2$ heating rate, which we find to be $1 / (2.2\,\text{ms})$ \cite{supplement}. 

We convert this heating rate to a leakage rate per Clifford using the prescription in Ref.\,\cite{rbidles}. The resulting lower bound on leakage due to heating is shown in the dashed line in Fig.\,4(a). For pulses longer than 15\,ns, we find that the leakage rate is within a factor of 2 of this lower bound, confirming that even at relatively short timescales, we are being limited by incoherent processes. We note that the heating rate and $T_1$ decay rate are consistent with an equilibrium population of 0.8\% for the $|1\rangle$ state \cite{supplement}. In other works, equilibrium populations closer to 0.1\% have been achieved \cite{jin2014thermal}, suggesting that incoherent leakage can be reduced through improved thermalization.

In conclusion, we have used single qubit randomized benchmarking to study leakage errors in a superconducting qubit. Using RB, we show that simple DRAG correction alone cannot minimize leakage and total gate error simultaneously, but by detuning our pulses, we obtain gates with both high fidelity and low leakage. We also measured the dependence of leakage on pulse length, and found that heating of the qubit is a significant source of leakage in our system. Because RB is platform independent, this method should be applicable to other systems provided they have high fidelity measurement of their leakage states. This method should also be extendable to two-qubit gates, where entangling interactions can be a significant source of leakage \cite{martinis2014fast}. 

We thank Tobias Chasseur, Felix Motzoi, and Frank Wilhelm for valuable discussions. This work was supported by Google. Z. Chen and C. Quintana also acknowledge support from the National Science Foundation Graduate Research Fellowhship under Grant No. 
DGE 1144085. Devices were made at the UC Santa Barbara Nanofabrication Facility, a part of the NSF-funded National Nanotechnology Infrastructure Network, and at the NanoStructures Cleanroom Facility.
\bibliographystyle{apsrev}
\bibliography{GateError}

\begin{thebibliography}{29}
\bibitem[{foo()}]{footnote}
\bibinfo{note}{$T_1^{|1\rangle}$ is not $2 T_1^{|2\rangle}$ as might be
  expected because the transitions are at different frequencies, and $T_1$ in
  our system is highly frequency dependent \cite{barends2013coherent}.}
\end{thebibliography}
\end{document}


\title{Supplementary Information for ``Measuring and Suppressing Quantum State Leakage in a Superconducting Qubit''}

\author{Zijun Chen}
\affiliation{Department of Physics, University of California, Santa Barbara, California 93106-9530, USA}
\author{Julian Kelly}
\affiliation{Google Inc., Santa Barbara, California 93117, USA}
\author{Chris Quintana}
\affiliation{Department of Physics, University of California, Santa Barbara, California 93106-9530, USA}
\author{R. Barends}
\affiliation{Google Inc., Santa Barbara, California 93117, USA}
\author{B. Campbell}
\affiliation{Department of Physics, University of California, Santa Barbara, California 93106-9530, USA}
\author{Yu Chen}
\affiliation{Google Inc., Santa Barbara, California 93117, USA}
\author{B. Chiaro}
\affiliation{Department of Physics, University of California, Santa Barbara, California 93106-9530, USA}
\author{A. Dunsworth}
\affiliation{Department of Physics, University of California, Santa Barbara, California 93106-9530, USA}
\author{A. Fowler}
\affiliation{Google Inc., Santa Barbara, California 93117, USA}
\author{E. Lucero}
\affiliation{Google Inc., Santa Barbara, California 93117, USA}
\author{E. Jeffrey}
\affiliation{Google Inc., Santa Barbara, California 93117, USA}
\author{A. Megrant}
\affiliation{Department of Physics, University of California, Santa Barbara, California 93106-9530, USA}
\affiliation{Department of Materials, University of California, Santa Barbara, California 93106, USA}
\author{J. Mutus}
\affiliation{Google Inc., Santa Barbara, California 93117, USA}
\author{M. Neeley}
\affiliation{Google Inc., Santa Barbara, California 93117, USA}
\author{C. Neill}
\affiliation{Department of Physics, University of California, Santa Barbara, California 93106-9530, USA}
\author{P. J. J. O'Malley}
\affiliation{Department of Physics, University of California, Santa Barbara, California 93106-9530, USA}
\author{P. Roushan}
\affiliation{Google Inc., Santa Barbara, California 93117, USA}
\author{D. Sank}
\affiliation{Google Inc., Santa Barbara, California 93117, USA}
\author{A. Vainsencher}
\affiliation{Department of Physics, University of California, Santa Barbara, California 93106-9530, USA}
\author{J. Wenner}
\affiliation{Department of Physics, University of California, Santa Barbara, California 93106-9530, USA}
\author{T. C. White}
\affiliation{Department of Physics, University of California, Santa Barbara, California 93106-9530, USA}

\author{A. N. Korotkov}
\affiliation{Department of Electrical and Computer Engineering, University of California, Riverside, California 92521, USA}
\author{John M. Martinis}
\email{jmartinis@google.com}
\affiliation{Department of Physics, University of California, Santa Barbara, California 93106-9530, USA}
\affiliation{Google Inc., Santa Barbara, California 93117, USA}
\date{\today}
\begin{abstract}

\end{abstract}
\maketitle

\section*{Rate equation for $|2\rangle$ state population}

In this section we discuss the rate equation which describes the $|2\rangle$ state population in the randomized benchmarking (RB) procedure.

Neglecting the population of the $|3\rangle$ state and higher levels, it is natural to describe (phenomenologically) the {\it average} population $p_{|2\rangle}(m)$ of the state $|2\rangle$ after $m$ Cliffords using the evolution equation
\begin{equation}
p_{|2\rangle}(m+1) = p_{|2\rangle}(m) +\gamma_\uparrow [1-p_{|2\rangle}(m)] -  \gamma_\downarrow  p_{|2\rangle}(m), 
\label{rate-eq-1}
\end{equation}
where $\gamma_\uparrow$ is the probability of the $|2\rangle$ state excitation per Clifford, averaged over Cliffords and also over the initial state in the qubit subspace, while $\gamma_\downarrow$
is the probability of returning from the state $|2\rangle$ to the qubit subspace, averaged over Cliffords. We emphasize that Eq.\ (\ref{rate-eq-1}) would be invalid for a particular RB sequence, but we apply it only assuming averaging over the RB sequences: $p_{|2\rangle}(m)$, $\gamma_\uparrow$, and $\gamma_\downarrow$ are all the averaged values. So far we have introduced Eq.\ (\ref{rate-eq-1}) phenomenologically; we will discuss the applicability of this equation later. 

The solution to Eq.\, (\ref{rate-eq-1}) is 
\begin{equation}
p_{|2\rangle}(m) = C (1-\Gamma)^m +p_\infty, \,\,\,\, p_\infty=\frac{\gamma_\uparrow}{\Gamma}, \,\,\,\, \Gamma = \gamma_\uparrow + \gamma_\downarrow, 
\label{evol-1}
\end{equation} 
where $C$ is a constant, determined by the initial condition, $C=p_{|2\rangle}(0)-p_\infty$. In the case $\Gamma \ll 1$ this solution can be replaced with
\begin{equation}
p_{|2\rangle}(m) = [p_{|2\rangle}(0)-p_\infty ] \, e^{-\Gamma m} + p_\infty ,
\label{evol-2}
\end{equation} 
that obviously corresponds to the standard rate equation 
\begin{equation}
    dp_{|2\rangle}(m)/dm = \gamma_\uparrow [1-p_{|2\rangle}(m)] -  \gamma_\downarrow  p_{|2\rangle}(m),
\label{rate-eq-2}
\end{equation}

to which Eq.\, (\ref{rate-eq-1}) reduces when $m$ is considered as a quasicontinuous variable ($m\gg 1$). Thus, $m$ plays the role of the dimensionless time, while $\gamma_\uparrow$ and $\gamma_\downarrow$ are the excitation and relaxation rates in this dimensionless time.
Note that if $p_{|2\rangle}(0)=0$, then Eq.\, (\ref{evol-2}) becomes $p_{|2\rangle}(m) = p_\infty (1- e^{-\Gamma m})$. 

Also note that if observed probabilities $\tilde p_{|2\rangle}$ are different from actual probabilities $p_{|2\rangle}$ due to measurement infidelity in a linear way, $\tilde p_{|2\rangle}(m)=A p_{|2\rangle}(m) +B [1-p_{|2\rangle}(m)]$ (here $A\approx 1$ is the fidelity of the state $|2\rangle$ measurement, while $B\ll 1$ is the average probability of misidentifying a state within the qubit subspace as the $|2\rangle$ state), then Eqs.\ (\ref{rate-eq-1})--(\ref{rate-eq-2}) remain valid for  $\tilde p_{|2\rangle}(m)$, but with the slightly changed rates: $\gamma_\uparrow \rightarrow \tilde\gamma_\uparrow=A\gamma_\uparrow + B \gamma_\downarrow$, $\gamma_\downarrow \rightarrow \tilde\gamma_\downarrow =\Gamma -\tilde\gamma_\uparrow$,  $\tilde\Gamma =\Gamma$. Therefore, the rates $\tilde\gamma_\uparrow$ and $\tilde\gamma_\downarrow$ extracted from the RB results, may slightly differ from the actual rates $\gamma_\uparrow$ and $\gamma_\downarrow$.

Next we discuss the applicability of the rate equation (\ref{rate-eq-1}) for the  $|2\rangle$ state population. A rate equation usually assumes incoherent processes. However, in our case both coherent and incoherent processes are important: while the rate $\gamma_\downarrow$ is mostly determined by incoherent energy relaxation, the rate $\gamma_\uparrow$ is mostly determined (at least for short gates) by a unitary evolution, though with possibly fluctuating pulse shapes. Therefore, it is not obvious if the simple rate equation is applicable. Note that we do not apply random $\pm 1$ pulses for the $|2\rangle$ state as was suggested \cite{epstein2014investigating,wallman2015robust,chasseur2015complete} for formal randomization of the coherent processes. In our opinion, for practical purposes it is not necessary because of different transition frequencies $\omega_{21}$ and $\omega_{10}$. To illustrate this argument, let us assume only coherent excitations of the $|2\rangle$ state and consider the evolution of the wavefunction $c_0|0\rangle+c_1|1\rangle+c_2|2\rangle$ in the rotating frame based on $\omega_{10}$. Then for a particular sequence of Cliffords (assuming $|c_2|^2 \ll 1$) 
\begin{equation}
    c_2(m)= c_2(0)+\sum\nolimits_k  g_{\uparrow,k} \, e^{-i(\omega_{21}-\omega_{10})t_k} , 
\label{c2(m)}
\end{equation}
where the complex number $g_{\uparrow , k}$ is the contribution from $k$th Clifford in the sequence ($\gamma_\uparrow = \langle |g_{\uparrow , k}|^2\rangle$) and $t_k$ is the start time of $k$th Clifford. For $\omega_{21}-\omega_{10} = 2\pi \times -212\, $MHz and elementary gate time $\agt 10\,$ns, it is unlikely that the phase shifts in Eq.\ (\ref{c2(m)}) are close to exact integers of $2\pi$. Therefore, even if averaging over Cliffords and initial states does not provide full randomization in the sense that $\langle g_{\uparrow , k}\rangle \neq 0$, the extra phase factor (accumulating with $k$) helps to average the contributions to zero, so that in this example $|c_2(m)|^2\propto m$ (from two-dimensional random walk), as would also be expected from a simple rate-equation model. Thus, we expect that the rate equation should work well for coherent contributions to the leakage, and since it also works for incoherent processes, we expect the rate equation to be well applicable to our RB procedure. Experimental results presented in the main text confirm this expectation.

\section*{Measurement Setup}
The measurement setup is largely as described in the supplementary information for Ref.\,\onlinecite{kelly2015state}, with two primary differences. First, the qubits are no longer statically biased with a programmable voltage source separate from the Z-control DAC. Instead they are operated by internally adding a DC offset to the output of the control DAC. As such, the bias tees and attenuators on the Z-control lines at the 20\,mK stage were removed. Second, the thermalization of all lines was improved by clamping the lines to all stages from 4\,K to 20\,mK using copper thermal anchors \cite{fang2015optimizing}.
\section*{State Discrimination}

\begin{figure}
\begin{centering}
\includegraphics[width=3.25 in]{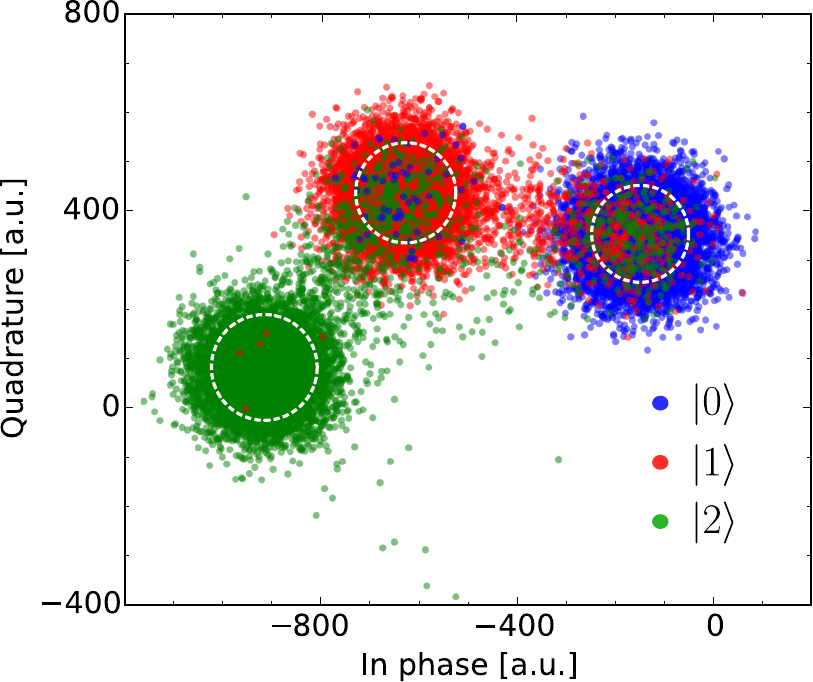} 
\par\end{centering}
\caption{Phase space points corresponding to the qubit being prepared in the $|0\rangle$ (blue), $|1\rangle$ (red), and $|2\rangle$ (green) states. Out of a total of 50,000 preparations of each state, 5000 are shown here. The states are discriminated based on their distance from the center of the cloud corresponding to each state. Points of one color positioned in a cloud of a different color indicate readout errors. The white circles in each cloud have radii corresponding to one standard deviation of the complex data in each cloud.}
\label{figure:readout} 
\end{figure}

Readout parameters for this device have previously been detailed in Ref.\,\onlinecite{kelly2015state}. At the operating point used for the experiment, we find the dispersive shift to be about 1\,MHz. We readout using a 1\,$\mu$s pulse. To characterize our readout fidelity, we prepare the qubit in each of the three states 50,000 times and measure. The raw IQ points of the demodulated signal \cite{jeffrey2014fast} are shown in Fig.\,S1. The probability of measuring the qubit in each state given preparation in another is as follows:
\[ \left(\begin{array}{ccc}
0.993 & 0.0069 & 5 \times 10^{-5} \\ 
0.055 & 0.945 & 5 \times {10^{-4}} \\ 
0.0246 & 0.083 & 0.892
\end{array}  \right)\]
where the row indicates the state prepared and the columns indicate the state measured. The primary source of error is $T_1$ decay of the excited states. The readout frequency was chosen to maximize the separation between the $|2\rangle$ state and the $|1\rangle$ state, resulting in a separation error between the two clouds of phase space points of around $1 \times 10^{-4}$.  However, the actual probability of preparing $|1\rangle$ and measuring $|2\rangle$ is greater, at around $5 \times 10^{-4}$. This is consistent with the heating rate of $4 \times 10^{-7}$ per nanosecond as measured in the main paper, multiplied by the readout time of 1\,$\mu$s. 

In general, we do not correct for measurement fidelity except in the thermalization measurement shown in Fig.\,4 of the main article. As noted above, the extraction of leakage rates from RB data is affected by readout fidelity. Thus, the leakage rates we quote are about 10\% lower than the actual leakage rates.

\section*{Dependence of Optimal Pulse Detuning on DRAG Weight and Pulse Length}
\begin{figure}
\begin{centering}
\includegraphics[width=3.25 in]{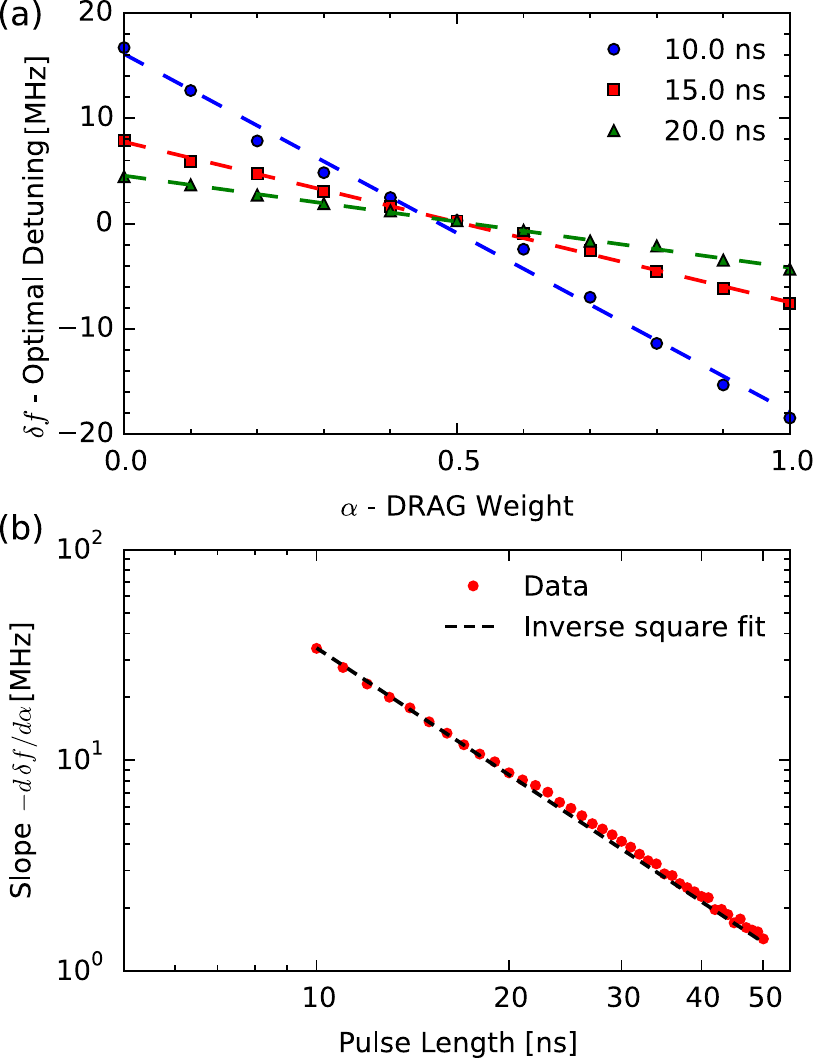} 
\par\end{centering}
\caption{(a) Dependence of the optimal detuning on $\alpha$. Three different pulse lengths are shown. The dashed lines are linear fits. (b) The slopes from the linear fits as shown in (a), for a range of pulse lengths. The dashed line is a fit to the inverse square of the pulse length, as expected from the AC Stark shift.}
\label{figure:detuneparams} 
\end{figure}
In Fig.\,2(a) we show the dependence of the optimal pulse detuning on the DRAG weight $\alpha$ for three different $\pi$-pulse lengths. For each pulse length, the dependence is linear, and the slope becomes more shallow with longer pulse length. In Fig.\,2(b), we plot the dependence of this slope on pulse length. We find that the slope between optimal detuning and DRAG is proportional to the inverse square of the pulse length. Equivalently stated, the slope depends quadratically on the drive strength, which we expect because the AC Stark shift scales quadratically with the strength of the driving field.

\section*{Leakage State Decay}

\begin{figure}
\begin{centering}
\includegraphics[width=3. in]{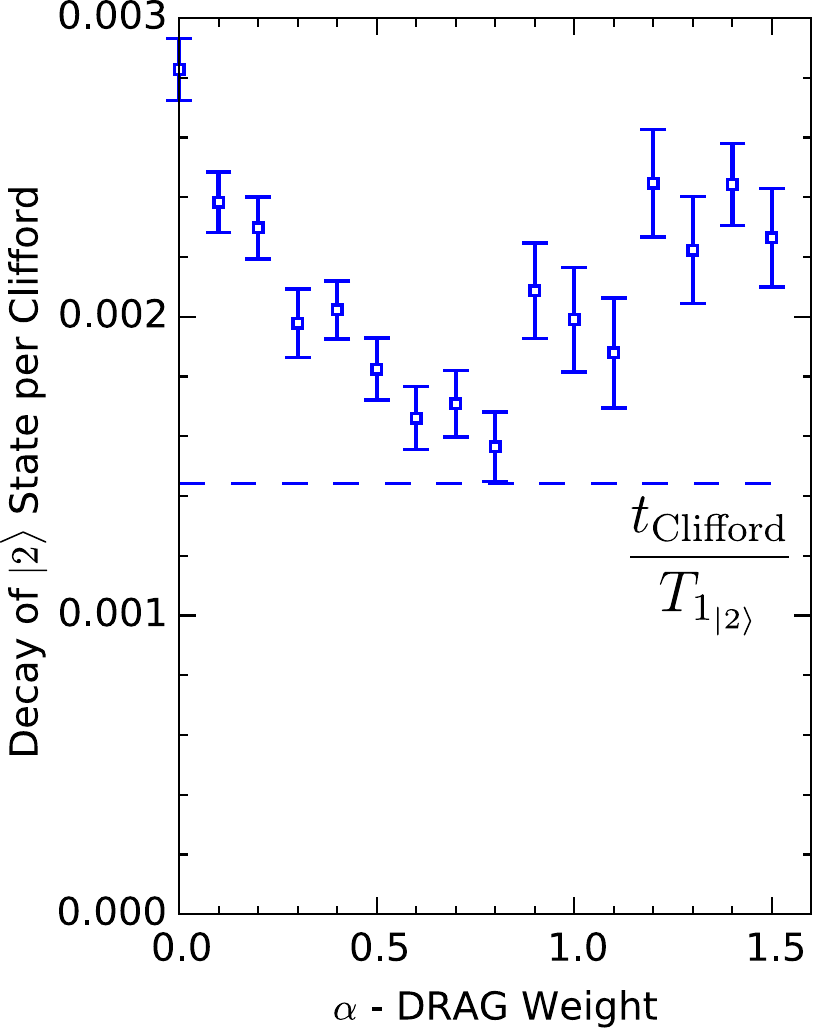} 
\par\end{centering}
\caption{Decay probability of the $|2\rangle$ state per Clifford measured using RB, corresponding to Fig.\,3(a) of the main paper. The dashed line indicates the expected incoherent decay from the measured $T_1$.}
\label{figure:readout} 
\end{figure}

Equation 2 contains both a leakage rate and a decay rate of the $|2\rangle$ state back into the computational subspace. We show in Fig.\,3 the decay rates corresponding to the data in Fig.\,3(a) of the main paper. The dashed line represents the decay expected due to $T_1$ decay of the $|2\rangle$ given an average Clifford time of $t_{\text{Clifford}}=18.75\,$ns. The $T_1$ for the $|2\rangle$ we use here is 13\,$\mu$s as measured concurrently with the RB data. We note that this is a different from the 18\,$\mu$s quoted in the context of Fig.\,4 of the main paper because these measurements were performed many days apart. Over that time scale, the fine features of the spectrum of two-level state (TLS) defects tend to drift. In general, the decay rates are higher than expected from $T_1$ decay. 

\section*{Raw Data for Simultaneously Optimized Fidelity and Leakage}
\begin{figure}
\begin{centering}
\includegraphics[width=3. in]{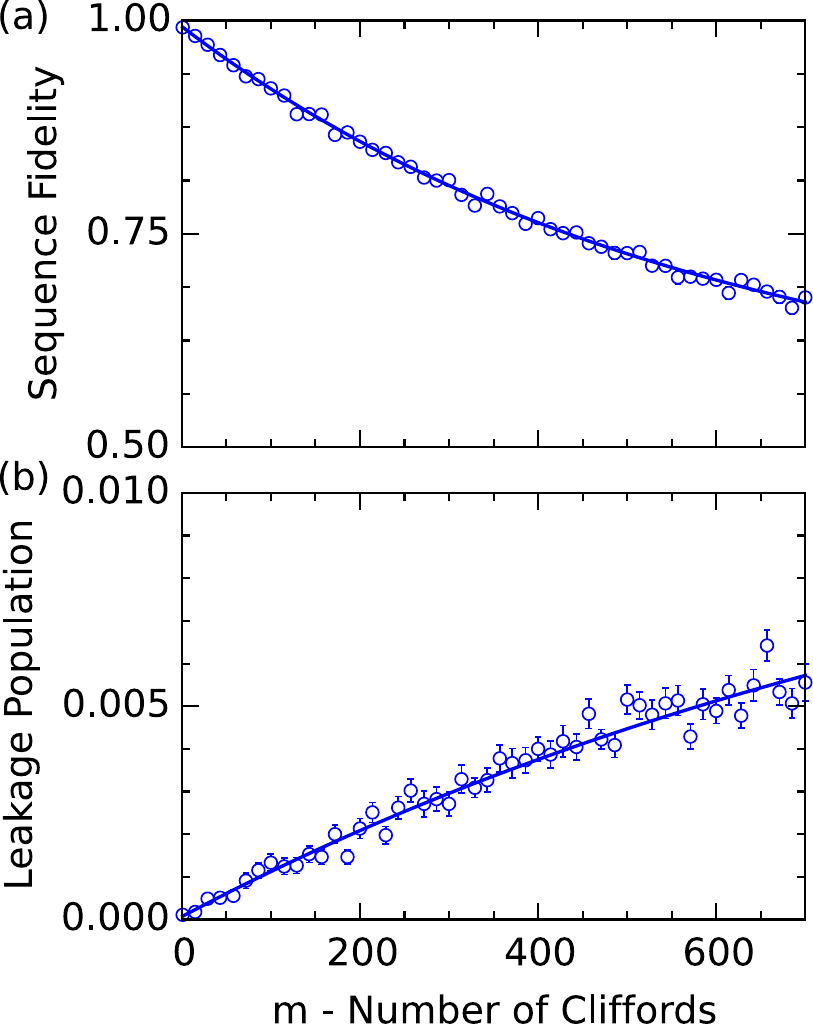} 
\par\end{centering}
\caption{Raw randomized benchmarking data for pulses optimized for both gate fidelity and leakage. (a) Sequence fidelity decay. The error per Clifford is $8.7 \pm 0.4 \times 10^{-4}$. (b) Leakage accumulation. The leakage per Clifford is $1.2 \pm 0.1 \times 10^{-5}$.}
\label{figure:readout} 
\end{figure}
In Fig.\,4, we show the raw randomized benchmarking data for 10\,ns pulses simultaneously optimized for fidelity and leakage, as described in Fig.\,3(b) of the main article. Here, $\alpha =$1.4, and  $\delta f = -30\,$MHz.

\section*{Thermalization at the $1 \leftrightarrow 2$ Transition Frequency}
\begin{figure}
\begin{centering}
\includegraphics[width=3.25 in]{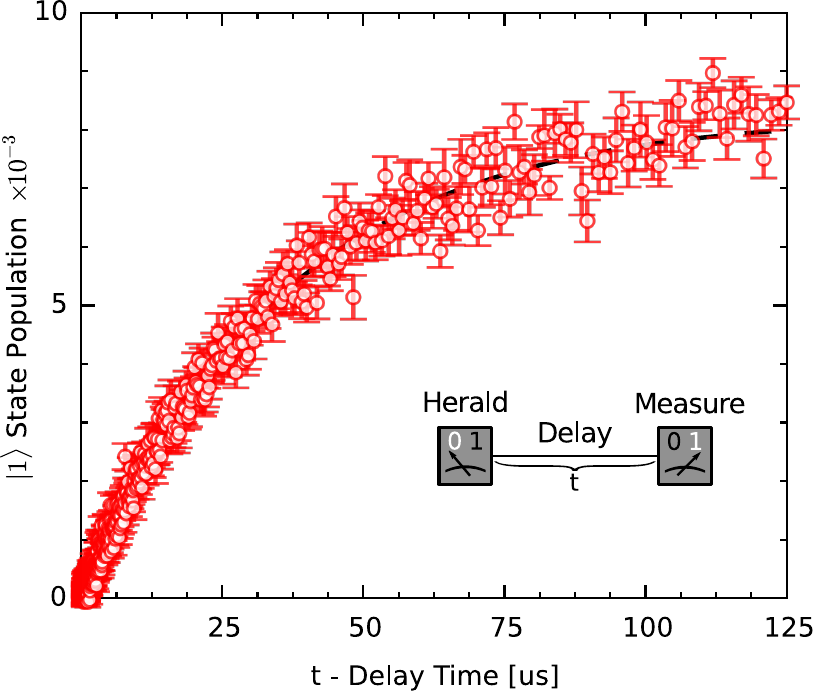} 
\par\end{centering}
\caption{Heating of the qubit, measured by heralding the $|0\rangle$ state, followed by a variable delay and a second measurement. The dashed line is a fit to a rate equation, where the only free parameter is the heating rate.}
\label{figure:readout} 
\end{figure}
To verify the heating rate measured in Fig.\,4 of the main article, we bias the qubit so that the 0$\leftrightarrow$1 transition frequency is equal to the original 1$\leftrightarrow$2 frequency, which was about 5.1\,GHz. We measure the $T_1$ of the $|1\rangle$ state at this frequency to be 39\,$\mu$s, roughly a factor of two greater than the measured $|2\rangle$ state $T_1$ of 18\,$\mu$s, as expected \cite{peterer2015coherence}. Next, we measure the heating rate of the 0$\leftrightarrow$1 transition by performing two measurements separated by a variable delay time, as shown in Fig.\,5. The first measurement heralds the $|0\rangle$ state to ensure the qubit is in $|0\rangle$ at $t=0$, and the second measurement probes the approach of the qubit to the equilibrium population. We fit to a rate equation with two rates, the heating rate and the $T_1$ decay rate; with the $T_1$ fixed by the previous measurement, we fit the heating rate to be $1/(4.7\,\text{ms})$. Again, we find the heating time constant to be roughly a factor of 2 larger than that of the $|2\rangle$ state, which we measured to be 2.2\,ms.

\section*{DRAG with Second Derivative Correction}
\begin{figure}
\begin{centering}
\includegraphics[width=3.25 in]{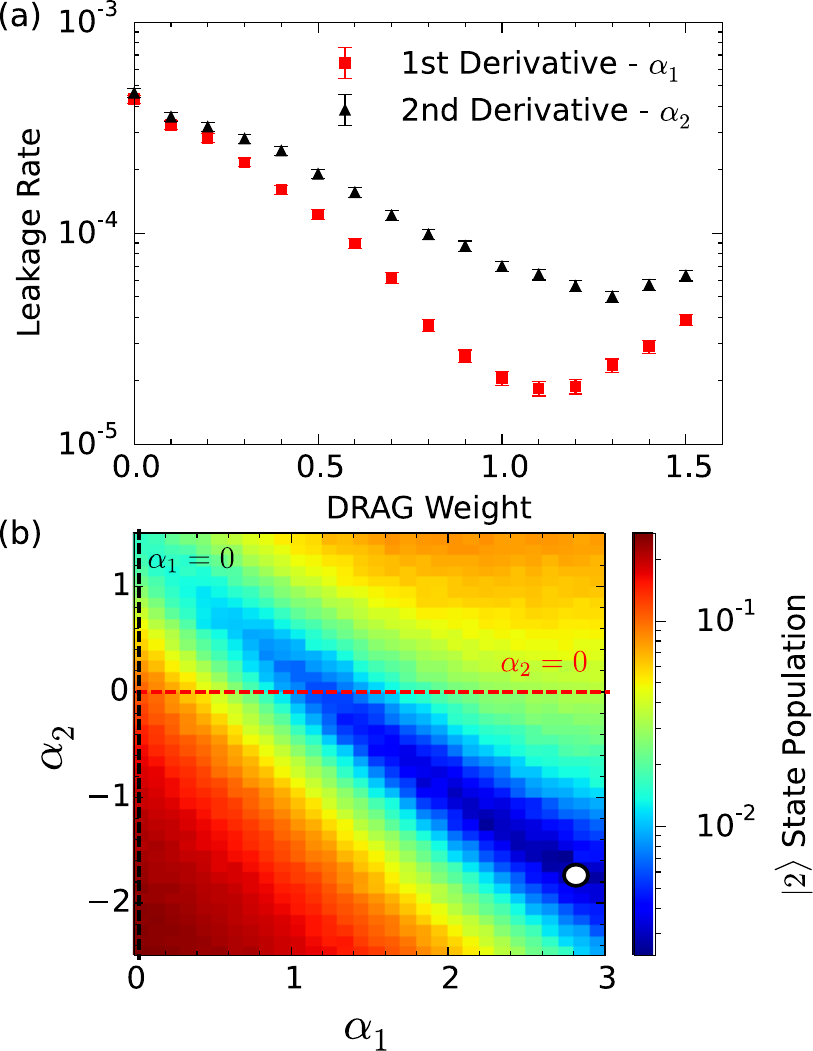} 
\par\end{centering}
\caption{Suppressing leakage using second derivative DRAG. (a) Leakage rate extracted from full Clifford based RB {\it vs} DRAG weighting ($\alpha_1$ and $\alpha_2$), using first derivative correction (red) and second derivative correction (black). Data is for 10\,ns pulses. (b) Leakage performance when using both first and second derivative DRAG. The color corresponds to the $|2\rangle$ state population after 700 Cliffords, and is the average of 45 different random sequences. The scale of the color is logarithmic. The dashed, horizontal red line corresponds to first derivative correction only while the vertical black line corresponds to second derivative correction only. The open circle highlights the minimum leakage population, which was $3 \times 10^{-3}$.}
\label{figure:readout} 
\end{figure}
Reference \onlinecite{motzoi2013improving} notes that for long pulses and large anharmonicity,  leakage can be suppressed using a DRAG-like technique with higher order derivatives. For example, DRAG correction with the second derivative takes the following form:

\begin{equation}
\Omega'(t) = \Omega(t) + \frac{\alpha_2}{\Delta^2} \ddot{\Omega}(t)
\end{equation}
where $\alpha_2$ is a weighting parameter. Note that unlike DRAG with first derivatives, the second derivative correction is applied to the in-phase component, which means that it does not have any effect on phase errors. We perform the same experiment as in Fig.\,3(a) of the main paper to compare first and second derivative DRAG correction for 10\,ns pulses without any detunings. As seen in Fig.\,6(a), the second derivative correction does indeed suppress  leakage, with a minimum leakage rate of $5 \times 10^{-5}$ at $\alpha_2 = 1.3$. However, the first derivative correction is still more effective by about a factor of 3 when optimized. Next, we implement both first and second derivative corrections simultaneously.
\begin{equation}
\Omega''(t) = \Omega(t) -i\frac{\alpha_1}{\Delta} \dot{\Omega}(t)+\frac{\alpha_2}{\Delta^2} \ddot{\Omega}(t)
\end{equation}
Because we have increased the dimension of our parameter space, performing full RB characterization by measuring leakage population versus sequence length for each set of parameters would take a prohibitively long time. Instead, we measure the leakage population for many random sequences but only for a single, large sequence length. We aim to measure the leakage state population near saturation, which is correlated with the leakage rate if the decay rate of the $|2\rangle$ state is mostly independent of the parameters under consideration. In Fig.\,6(b), we show the $|2\rangle$ state population after 700 Clifford gates, averaged over 45 different random sequences, while varying both the first and second derivative DRAG weights. We see that there is a substantial parameter space over which leakage can be suppressed. We obtain a minimum leakage population after 700 Cliffords of $3 \times 10^{-3}$ for $\alpha_1=2.8$ and $\alpha_2=-1.8$, which is a factor of 2 improvement over using only first derivative correction (e.g. as seen in Fig.\,4). However, using such a large $\alpha$ would also require a large detuning to compensate for phase errors, which will increase leakage. Thus, while our data suggests that there are still gains to be made in leakage performance, simultaneously optimizing for fidelity and leakage while using second derivative DRAG is non-trivial and an ongoing topic of research.
\bibliographystyle{apsrev}
\bibliography{supplement}